\documentstyle[12pt,aaspp4]{article}

\begin{document}

\title{A two zone model for the broad Iron line emission in MCG-6-30-15 }

\author{\bf R. Misra}
\affil{Inter-University Centre for Astronomy and Astrophysics, Pune, India}
\authoremail{rmisra@iucaa.ernet.in}

\begin{abstract}
We reanalyze the ASCA and BeppoSAX data of MCG-6-30-15, using a 
double zone model for the Iron line profile. In this model, the
X-ray source is located around $\approx 10$ Schwarzschild radius and
the regions interior and exterior to the X-ray source produce the
line emission. We find that this model fits the data with similar
reduced $\chi^2$
as the standard single zone model. The best fit inclination angle  
of the source ($i \approx 10^o$) for the medium intensity ASCA data set 
is compatible with that determined
by earlier modeling of optical lines. The observed
variability of the line profile with intensity can be explained 
as variations of the X-ray source size. That several AGN with
broad lines have the peak centroid near $6.4$ keV can be explained
within the framework of this model under certain conditions.
 
We also show that the simultaneous broad band observations of this
source by BeppoSAX rules out the Comptonization model which was
an alternative to the standard inner disk one. We thereby
strengthen the case that the line broadening occurs due to
the strong gravitational influence of a Black Hole.
\end{abstract}


\section{Introduction}

A long duration (4.2 days), observation of the Seyfert 1 AGN, 
MCG-6-30-15, revealed for the first time that the Iron line profile
in this source is broad (Tanaka et al. 1995). Subsequently
broad Iron lines were also detected in other AGN by ASCA 
( Nandra et al. 1997). Independent reconfirmation of this result
came from the broad band observations of this source by
BeppoSAX (Guainazzi et al. 1999). Recently a second long observation
of MCG-6-30-15 by ASCA confirmed the results obtained earlier (Iwasawa et al. 1999).

Tanaka et al. (1995) pointed out that the large width of the line 
could probably be due
to the extreme gravitational effects near the vicinity of a Black hole
(Fabian et al. 1989). In this model, the emission arises from
the innermost region ($\approx 6 - 10 r_g$ where $r_g = GM/c^2$) 
of a cold accretion disk around a Black hole which
is illuminated by an X-ray source.
Fabian et al. (1995) considered several  mechanisms for line
broadening 
and concluded that none of them were satisfactory except the 
inner disk emission model. One
of the alternate models considered by them is the Comptonization
model first proposed by Czerny, Zbyszewska and Raine (1991). In
the Comptonization model, line broadening occurs because of 
Compton down scattering of the emission line photons as they
pass through an optically thick cloud. Misra \& Kembhavi (1999)
showed that a compact highly ionized cloud was not ruled out
by the data available then. Subsequently, Misra \& Sutaria (1999)
fitted the ASCA data to the Comptonization model and found that
the fit was as good as the disk emission model. However,
It was pointed
out by Misra \& Kembhavi (1999) that 
simultaneous broad band data would be able to rule out or 
confirm the presence of such a cloud. The recent BeppoSAX
broad band ( 0.1-200 keV) observations of this source gives 
such an opportunity. In this paper, we show that analysis of
the BeppoSAX data indeed rules out the Comptonization model,
thereby strengthening the case for the disk line model.

Despite the success of the inner disk model, the actual
geometry of the source especially the position of the illuminating
X-ray source is still unknown. The inner disk emission 
models (Fabian et al. 1989; Laor 1991) 
used for fitting the data, assumes a cold accretion disk
inclined at a angle ($i$) with a power-law type radial 
emissivity function ($I \propto R^{-\alpha}$). Here $\alpha$
is called the emissivity index. Other parameters are the
inner ($R_{i}$) and outer ($R_{o}$) disk radius. Spectral
fitting of the ASCA data reveals that $\alpha$ is positive (Iwasawa 1996)
which implies that the X-ray source must be located near or at a radius
less than the inner edge of the disk ($6 - 1.2 r_g$). 
This is contrary to standard models
for the X-ray production in which the source is located where the maximum
gravitational energy is dissipated ($\approx 10 r_g$). In particular,
the hot disk model, in which the X-rays are generated in an inner hot
region of disk (Shapiro, Lightman \& Eardley 1976) is ruled out. The X-ray
source may then be in the form of a 
hot corona on top of the cold disk (Liang \& Price 1977). Even, here the
Iron line modeling restricts the location of the corona to be only
over the inner edge of the disk and not around $10 r_g$. 
It is important to study the constrains imposed by the Iron line
profile fitting in detail, since it may significantly change the
standard X-ray production models or force introduction of
relatively new ones ( for e.g. the X-ray source located in
the form of a jet).

The disk line fit to MCG-6-30-15 to the ASCA data constrains the
inclination angle $i \approx 30^o$ (Iwasawa et al. 1996). This
is contrary to the inclination
angle derived by modeling the optical H$\alpha$
measurements for the same source (Sulentic et al. 1998). Recently
Rokaki \& Boisson (1999) developed an accretion disk model for 
both the UV continuum and the optical lines from AGN. They also found
that for MCG-6-30-15, the best fit inclination angle $i \approx 12^o$. This
discrepancy could be due to a) the optical line are not produced
in the outer regions of the accretion disk, b) the disk is warped
or c) that the disk line model is simplistic i.e. the geometry assumed
is not a good approximation. That the geometry of the disk line model
is complex is also indicated by the study of several AGN with broad
line which revealed that many of them
have their peak centroid close to $6.4$ keV. 
It was pointed out by Sulentic, Marziani \& Calvani (1998) that 
this is not expected if the disks are
oriented in random directions. 
Taking these arguments into account
Sulentic, Marziani \& Calvani (1998) suggested that perhaps the
line profile is a sum of two independent components. Alternatively,
Blackman (1999) suggested that
this could be due to a concave inner accretion disk.

In this paper, we reanalyze the ASCA and BeppoSAX observation
of MCG-6-30-15 to study whether the Iron line profile could
really be a sum of two components and whether the line profile
modeling can be made more compatible with the standard X-ray
production models. In particular, if the X-ray source is located
near $\approx 10 r_g$, there will be two distinct regions for
line production. One will be the inner most region with radii
less than the X-ray producing source. Here the emissivity index ($\alpha$)
would be negative. There would also be a second region with
radii greater than the X-ray source where $\alpha$ would be positive.
We allow for the possibility that near the X-ray source, line emission
may not arise due to the absence of cold accretion disk or if the
disk is highly ionized.

\section{Results}

ASCA observed MCG-6-30-15 for $\approx 4.2$ days from 
1994 July 23 to 27 (Tanaka, Inoue \& Holt 1994). This long exposure
allowed for the first time the 
detection of a broad Iron line (Tanaka et al. 1995).
A subsequent detailed analysis by Iwasawa  et al. (1996) showed that
the line shape was variable and correlated to the intensity of the source.
Iwasawa et al. (1996) grouped the data into three intensity levels called
the low (LI), medium (MI) and high intensity (HI) data sets. During most 
($3/4$) of the observation time, the source was in the medium 
intensity level.
Iwasawa et al. fitted the line profile with a phenomenological two Gaussian
model and  accretion disk models around stationary (Schwarzschild)
and rotating (Kerr) Black holes. This data has also been
reanalyzed using alternate models like a Comptonizing cloud model
(Misra \& Sutaria 1999) and an occultation model (Weaver \& Yaqoob 1998).

Following Iwasawa et al. (1996) we divide the data into the three
different intensity levels. We analyze
the $3 - 10$ keV data since  below $3$ keV the 
spectrum is affected by the partially ionized gas (``the warm absorber'') 
surrounding the source. Data from both the SIS chips (SIS 1 and 2) for the 
Bright and Bright2 modes
were grouped and analyzed together.
Only the relative normalization between
these four sets of data  was allowed to vary, but in all cases the variation 
was found to be less than 2\%.

In table 1, we summarize the results for fitting the medium intensity (MI) 
data set. We start with fitting the data with the standard single zone
accretion disk model ( xspec model ``diskline'' , Fabian et al. 1989),
which through out this paper we refer to as the standard disk model.
To be consistent with the broad band BeppoSAX results ( Guainazzi et al. 1999)
we include the possibility of an Iron edge in the spectrum with threshold
energy fixed at $7.6$ keV. The inner edge of the accretion disk is
fixed at 6 $r_g$. We also fit the data with a nearly maximally
rotating Black hole model ( xspec model ``laor'' , Laor  1991) 
with the inner edge fixed at 1.2 $r_g$. Both the models fit the
data equally well with  $\Delta \chi^2 = 2$ for $1421$ degrees
of freedom between them. These results are consistent with those
reported by Iwasawa et al (1996).
An additional Gaussian line emission is not required
for this model. The angle of inclination ($i$) is well constrained
for both the fits to be $\approx 30^o$. So fixing the inclination 
angle to be $10^o$ gives a unacceptable increase in $\Delta \chi^2 = 82$
(Table 1: row 3). However, an addition of a narrow Gaussian line with
centroid energy of $6.4$ keV reduces $\chi^2$ by $52$ (Table 1: column 4).
Thus if the inclination angle for the source is restricted to
be around $10^o$ by some other observational constrain (e.g. modeling of
the  optical lines by Sulentic et al. 1998), then an additional component 
will be required to
explain the ASCA observations considered here. If $i$ is fixed at
$20^o$ a similar result is obtained ( reduced $\chi^2 = 1420$ for
1421 degrees of freedom), however the strength and significance
of the narrow Gaussian line is slightly reduced ($I_g = 3 \times 10^{-5}$
photons sec$^{-1}$ cm$^{-3}$). The reduced $\chi^2$ for the model
with a constrained inclination angle is still significantly higher
than the standard disk model (Table 1: column 1).

\begin{table}
\caption{Spectral Parameters for the ASCA medium intensity data set. Parameters without errors were fixed during fitting.}
\begin{tabular}{ccccccc}
\hline
 Model & Units & &&&&\\
param. & && &&&\\
\hline
\hline
$E_{th}$ & keV & 7.6 &7.6 & 7.6 & 7.6 & 7.6\\
$\tau$ &  & $0.09^{+0.06}_{-0.06}$& $0.04^{+0.09}_{-0.04}$ & $0.22^{+0.06}_{-0.06}$ & $0.14^{+0.06}_{-0.06}$& $0.11^{+0.06}_{-0.06}$\\
$N_H$ & $10^{20}$ cm$^{-2}$  & 6.4 & 6.4& 6.4& 6.4 & 6.4\\
\hline
$\Gamma$ &  & $2.01^{+0.03}_{-0.03}$ & $2.04^{+0.05}_{-0.05}$ & $1.91^{+0.02}_{-0.03}$ & $1.97^{+0.03}_{-0.03}$ & $2.00^{+0.03}_{-0.03}$\\
\hline
i & deg  & $31.7^{+1.0}_{-1.0}$ & $32.1^{+1.8}_{-1.6}$ & 10 & 10& $11.4^{+1.4}_{-1.2}$\\
\hline
$E_{d1}$ & keV & 6.4 & 6.4 &6.4 & 6.4 & 6.7\\
$\alpha_1$ &  & $1.25^{+0.75}_{-0.83}$ & $1.60^{+0.6}_{-0.7}$ &$2.86^{+0.19}_{-0.14}$& $3.46^{+0.35}_{-0.31}$ &  -3.0\\
$R_{i1}$ & $r_g$  & 6.0 & 1.2 & 6.0& 6.0 & 6.0\\
$R_{o1}$ & $r_g$  & $17^{+2}_{-1}$ & $17.6^{+3.8}_{-1.9}$ & 1000 & $33^{+15}_{-8}$ & $8.6^{+0.5}_{-0.8}$\\
$I_{d1}$ & $10^{-5}$ s$^{-1}$ cm$^{-2}$  & $15.9^{+1.2}_{-1.4}$ & $20.0^{+4.0}_{-3.5}$ & $8.2^{+1.0}_{-1.1}$ & $8.6^{+1.1}_{-1.0}$& $6.5^{+0.8}_{-0.9}$\\
\hline
$E_{d2}$ & keV & - & -&- & - & 6.7\\
$\alpha_2$ & & - & - & - & - & 3.0\\
$R_{i2}$ & $r_g$ & - & - & - & - & $12.7^{+1.5}_{-1.3}$\\
$R_{o2}$ & $r_g$ & - & - & - & - & 1000\\
$I_{d2}$ & $10^{-5}$  s$^{-1}$ cm$^{-2}$ & - & - &- & - & $7.8^{+0.9}_{-0.8}$\\
\hline
$I_g$ & $10^{-5}$  s$^{-1}$ cm$^{-2}$  &  - & - & - & $3.9^{+0.5}_{-0.6}$& -\\
\hline
$\chi^2$/(dof) &   & 1381(1421) & 1379 (1421) & 1463(1422) & 1411(1421) & 1380 (1420)\\
\hline
\hline
\end{tabular}
\end{table}

An additional component in the line shape, if present, would
also be expected to arise from close to the Black hole, since
this feature is also variable in short time scales ($<$ hrs). 
There could be two regions in the inner accretion disk which
produces the Iron line. Such a double zone region is expected
if the X-ray source is located near the radius of maximum 
gravitational energy dissipation ($\approx 10 r_g$).
The picture would then be of a cold accretion disk with an X-ray source
in the form of an extended corona around $10 r_g$. Alternately, the
X-ray producing region could be a hot accretion disk around $10 r_g$ with 
a cold outer disk and an inner region which is again cold. In either
case there
would be an inner most line emitting region ($6 r_g < R < 10 r_g$) 
where the emissivity index ($\alpha$ where $I \propto R^{-\alpha}$)
would typically be positive. A outer second region ($10 r_g  < R$) would
also be impinged by the X-rays but in this case $\alpha$ would
be expected to be negative. The two regions will be
distinct if the X-ray source is an hot disk in between the two.
On the other hand, if the X-ray source is in the form of a corona,
the underlying cold disk may be highly ionized and line emission
may not arise from around $10 r_g$. We approximate the complex
geometry above by a simple model consisting of two standard disk
line models ( i.e. two xspec model: ``diskline''). For the
first diskline model the inner radius ($R_{i1}$) is held constant
at $6 r_g$ and the emissivity index ($\alpha_1$) is fixed at a
constant negative value ( $ = -3$). The outer radius ($R_{o1}$) is
a free parameter. For the second diskline model the outer radius ($R_{o2}$)
is held constant at a large value ($ = 1000 r_g$), while the inner
radius $(R_{i2})$ is a free parameter. The emissivity index ($\alpha_2$) 
is fixed at a constant positive value ( $ = 3$). The results obtained
here are not very sensitive to the actual values of ($\alpha$) chosen
if $\alpha_1$ is taken to be negative and $\alpha_2$ is taken to be positive.
In this phenomenological description, the region of the disk between
($R_{o1} < r < R_{i2}$) is either the X-ray producing hot disk or
where the cold disk is highly ionized. This model will be refered to
in this paper as the double zone model. Table 1: column 5 shows the
result of fitting such a model to the ASCA MI data set. The reduced
$\chi^2$ is similar to the one obtained using the standard disk
model. The best fit inclination angle ($i \approx 10^o$) agrees
well with the value obtained on modeling the optical line emission 
from the source ( Sulentic et al. 1988).  A better fit to the
spectrum is obtained if the rest frame
energy of the line emission from both the regions is $6.7$ keV. This
indicates that the Iron in the cold disk is partially ionized. The
line profile for the double zone model is shown in figure 1. We
note that the combined line profile can be complex with several
features. In simple disk line fits, these features may be
interpreted as an absorption edge around $\approx 6$ keV as
was done for another AGN, NGC 3516 by Nandra et al. (1999). 

Table 2 summarizes the results obtained by fitting the above models
to the ASCA high intensity (HI) data set. Since the statistics for
this data set is lower than for the MI data set, some parameters
($\tau$, $i$, $\alpha$ and $R_{o1}$) have been fixed. As pointed
out by Iwasawa et. al. (1996) the HI data set is not well described
by the standard disk line model and an additional narrow line Gaussian
is required (reduced $\chi^2$ decreases by 20). The double zone model
also fits the data better than the standard disk line one 
(reduced $\chi^2$ decreases by 14).

\begin{table}
\caption{Spectral Parameters for the ASCA high intensity data set.
 Parameters without errors were fixed during fitting.}
\begin{tabular}{cccccc}
\hline
 Model & Units & &&&\\
parameters & & &&&\\
\hline
\hline
$E_{th}$ & keV & 7.6 &7.6 & 7.6 & 7.6 \\
$\tau$ &  & 0.1 & 0.1 & 0.1 & 0.1\\
$N_H$ & $10^{20}$ cm$^{-2}$  & 6.4 & 6.4 & 6.4& 6.4 \\
\hline
$\Gamma$ &  & $1.96^{+0.07}_{-0.03}$ & $1.98^{+0.08}_{-0.08}$ &  $1.92^{+0.08}_{-0.03}$ & $1.98^{+.04}_{-.05}$\\
\hline
i & deg  & 30 & 30 & 10 & 10\\
\hline
$E_{d1}$ & keV & 6.4 & 6.4 & 6.4 & 6.7 \\
$\alpha_1$ &  & 2.0 & 2.0 & 3.5& -3.0 \\
$R_{i1}$ & $r_g$  & 6.0 & 1.23 & 6.0& 6.0 \\
$R_{o1}$ & $r_g$  & $28.5^{+16}_{-10.2}$ & $15.1^{+0.9}_{-0.9}$ & 35 & 8.5 \\
$I_{d1}$ & $10^{-5}$ ph s$^{-1}$ cm$^{-2}$  & $11.8^{+5.4}_{-3.0}$ & $12.1^{+4.7}_{-2.5}$ &$1.7^{+3.4}_{-1.7}$ & $4.0^{+2.3}_{-2.0}$\\
\hline
$E_{d2}$ & keV & - & - & - & 6.7 \\
$\alpha_2$ & & - & - & - & 3.0 \\
$R_{i2}$ & $r_g$ & - & - & - & $17.5^{+5.0}_{-3.5}$ \\
$R_{o2}$ & $r_g$ & - & - & - & 1000 \\
$I_{d2}$ & $10^{-5}$ ph s$^{-1}$ cm$^{-2}$ & - & - & - & $8.6^{+1.4}_{-1.4}$ \\
\hline
$I_g$ & $10^{-5}$ ph s$^{-1}$ cm$^{-2}$  &  - & - & $6.3^{+1.5}_{-1.5}$ & -\\
\hline
$\chi^2$/(dof) &   & 506(497) & 500 (497) & 486(496) & 491(496) \\
\hline
\hline
\end{tabular}
\end{table}

Table 3  summarizes the results obtained by fitting the above models
to the ASCA low intensity (LI) data set. In this intensity level
the source exhibits an extended red wing which has been interpreted
as emission from a  disk that extends to less than $6 r_g$ which would
imply that the Black hole is spinning at a nearly maximal rate 
(Iwasawa et al. 1996). Here also we find that the $\Delta \chi^2 
= 4$ (for 153 dof) between the two models ( Table 3: column 1 and 2).
However, as pointed out by Weaver \& Yaqoob (1998) although
this difference is statistically significant the inclusion of
systematic errors may decrease its overall significance. Moreover,
the spectral calculations of this models are approximate and
for simple geometries. It should be noted that for $R = 6 r_g$ 
and inclination angle $i = 0^o$, the
red-shift, $E_{obs}/E_{emit} = (1.-3r_g/r)^{1/2} = 0.707$
which means $E_{obs} = 4.5$ keV for $6.4$ keV rest frame photons
(Hanawa 1989). A similar exercise for
$i = 90^o$, shows that the maximum $E_{obs} =
9$ KeV and the minimum $E_{obs} = 3$ keV. Hence just
the detection of photons at 4 keV does not necessarily
imply that the metric is Kerr. The spectral fit for
Kerr is better than the Schwarzschild case because of
the difference in spectral shape. 
For the double zone model, if both $R_{i2}$ and $R_{o1}$ are allowed to
be  free parameters then the unphysical result of $R_{o1} > R_{i2}$
is obtained. Hence, for this data set we have imposed the
condition $R_{o1} = R_{i2}$. The reduced $\chi^2$ obtained is
similar to the standard disk model (columns 1 and 4). If
the rest frame energy of the inner region is $6.4$ keV 
the reduced $\chi^2$ obtained is
similar to the rotating Black hole case (columns 2 and 5).
Since the X-ray luminosity is lower in this data set, the
inner disk may indeed be at a lower ionized state.
The
line profile for the double zone model is shown in figure 2.

\begin{table}
\caption{Spectral Parameters for the ASCA low intensity data set.
 Parameters without errors were fixed during fitting.}
\begin{tabular}{ccccccc}
\hline
 Model & Units & &&&&\\
parameters & & &&&&\\
\hline
\hline
$E_{th}$ & keV & 7.6& 7.6 & 7.6 & 7.6 & 7.6\\
$\tau$ &  & 0.1 & 0.1 &0.1 & 0.1& 0.1\\
$N_H$ & $10^{20}$ cm$^{-2}$  & 6.4 & 6.4 &6.4& 6.4 & 6.4\\
\hline
$\Gamma$ &  & $1.78^{+0.14}_{-0.11}$ & $1.81^{+0.15}_{-0.12}$ & $1.77^{+0.12}_{-0.12}$ & $1.78^{+0.14}_{-0.11}$ & $1.81^{+0.12}_{-0.12}$\\
\hline
i & deg  & 30 & 30 & 10 & 10& 10\\
\hline
$E_{d1}$ & keV & 6.4 & 6.4 & 6.4 & 6.7 & 6.4\\
$\alpha_1$ &  & $1.1^{+0.5}_{-5.3}$ & $2.1^{+0.7}_{-1.2}$ & 3.5& -3.0 &  -3.0\\
$R_{i1}$ & $r_g$  & 6.0 & 1.23 & 6.0& 6.0 & 6.0\\
$R_{o1}$ & $r_g$  & 15.5 & 15.5 & 35 & $7.8^{+0.6}_{-0.6}$ & $9.1^{+0.6}_{-0.6}$\\
$I_{d1}$ & $10^{-5}$ ph s$^{-1}$ cm$^{-2}$  & $15.5^{+4.7}_{-5.0}$ & $21.4^{+7.5}_{-6.3}$ & $10.4^{+3.0}_{-2.4}$ & $5.5^{+2.1}_{-3.7}$& $6.7^{+1.9}_{-2.2}$\\
\hline
$E_{d2}$ & keV & - & - & - & 6.7 & 6.7\\
$\alpha_2$ & & - & - & - & 3.0 & 3.0\\
$R_{i2}$ & $r_g$ & - & -  & - & 7.8 & 9.1\\
$R_{o2}$ & $r_g$ & - & - & - & 1000 & 1000\\
$I_{d2}$ & $10^{-5}$ ph s$^{-1}$ cm$^{-2}$ & - & - & - & $9.0^{+5.6}_{-2.6}$ & $9.1^{+5.3}_{-2.1}$\\
\hline
$I_g$ & $10^{-5}$ ph s$^{-1}$ cm$^{-2}$  &  - & - & $3.4^{+1.5}_{-1.4}$ & -& -\\
\hline
$\chi^2$/(dof) &   & 172(153) & 168 (153) & 172(153) & 171(152) & 168(152)\\
\hline
\hline
\end{tabular}
\end{table}

MCG-630-15 was observed by BeppoSAX from 1996 July 29 to
August 3 using the Low Energy Concentrator Spectrometer
(LECS, $0.1 - 4$ keV), the Medium  Energy Concentrator Spectrometer
(MECS, $1.8 - 10.5$ keV) and the Phoswitch Detector System
(PDS, $17 - 200$ keV) (Guainazzi et al. 1999). The low
energy spectrum is affected by the presence of ``a warm
absorber''. Following Guainazzi et al. (1999) and Orr
et al. (1997) we parameterized the warm absorber as
absorption edges with threshold energies: $0.74$ keV (O VII),
$0.87$ keV (O VIII) and  $1.2$ keV (Ne IX) and the following
emission lines with centroid energies: $0.62$ keV ($K_\alpha$ O VII)
and $0.86$ keV (iron-L). We also added an absorption edge with threshold
energy $7.6$ keV which corresponds approximately to Fe XV absorption.
For the continuum, an exponentially cutoff power-law with
a Compton reflection component (Xspec model: ``pexrav'',
Magdziarz \& Zdziarski 1995) was used.

\begin{table}
\caption{Spectral Parameters for the BeppoSAX data set.
 Parameters without errors were fixed during fitting.}
\begin{tabular}{cccccc}
\hline 
 Model & Units & &&&\\
parameters & & &&&\\
\hline
\hline
$E_{th}$ & keV & 7.6 & 7.6 & 7.6 & 7.6\\
$\tau$ &  & $0.14^{+0.03}_{-0.04}$ & $0.19^{+0.04}_{-0.03}$ & $0.17^{+0.03}_{-0.05}$& $0.16^{+0.03}_{-0.04}$\\
$N_H$ & $10^{20}$ cm$^{-2}$  & $6.4^{+0.2}_{-0.3}$ & $6.1^{+0.2}_{-0.3}$& $6.3^{+0.2}_{-0.3}$ & $6.2^{+0.3}_{-0.2}$\\
\hline
$\Gamma$ &  & $1.99^{+0.03}_{-0.04}$ & $1.94^{+0.03}_{-0.04}$ & $1.96^{+0.04}_{-0.04}$ & $1.97^{+0.04}_{-0.04}$\\
$E_{cut}$ & keV  & $108^{+57}_{-39}$ & $82^{+31}_{-20}$ & $91^{+40}_{-23}$ & $101^{+49}_{-27}$\\
$R_{refl}$ &  & $0.77^{+0.23}_{-0.21}$ & $0.67^{+0.23}_{-0.20}$ & $0.72^{+0.22}_{-0.21}$ & $0.65^{+0.22}_{-0.18}$\\
\hline
i & deg  & $37^{+2}_{-2}$ & 10 & 10& 10\\
\hline
$E_{d1}$ & keV & 6.4 & 6.4 & 6.4 & 6.7\\
$\alpha_1$ &  & 2.0 & $1.4^{+1.1}_{-3.0}$ & 3.5 &  -3.0\\
$R_{i1}$ & $r_g$  & 6.0 & 6.0& 6.0 & 6.0\\
$R_{o1}$ & $r_g$  & $7.5^{+2.4}_{-1.0}$ & 1000 & $8.0^{+6.8}_{-1.5}$ & $6.2^{+1.2}_{-0.2}$\\
$I_{d1}$ & $10^{-5}$ ph s$^{-1}$ cm$^{-2}$  & $9.5^{+1.7}_{-2.0}$ & $3.0^{+1.3}_{-0.9}$ & $2.6^{+1.3}_{-1.2}$& $2.8^{+3.4}_{-1.1}$\\
\hline
$E_{d2}$ & keV & - & - & - & 6.7\\
$\alpha_2$ & & - & - & - & 3.0\\
$R_{i2}$ & $r_g$ & - & - & - & $13^{+5}_{-4}$\\
$R_{o2}$ & $r_g$ & - & - & - & 1000\\
$I_{d2}$ & $10^{-5}$ ph s$^{-1}$ cm$^{-2}$ & - & - & - & $4.4^{+1.2}_{-0.9}$\\
\hline
$I_g$ & $10^{-5}$ ph s$^{-1}$ cm$^{-2}$  &  - & -& 3.5& -\\
\hline
$\chi^2$/(dof) &   & 120(123) & 132(123) & 123(123) & 124 (123)\\
\hline
\hline
\end{tabular}
\end{table}

Table 4 summarizes the result of fitting the various Iron line
models to the BeppoSAX data set. As reported by 
Guainazzi et al. (1999) the standard disk model fits
the data well. Like the ASCA results the inclination angle
is well constrained to be $i \approx 35^o$ and constraining
$i = 10^o$ requires an additional Gaussian line (Table 4:
column 2 and 3). The double zone model also fits the data
well but the inner line producing region is somewhat smaller
than the ASCA results ( $R_{o1} \approx 6.2 r_g$).

An alternate model to the disk line emission model is
the Comptonization model where the line is broadened
due Compton down scattering of the photons 
as they
pass through an optically thick cloud. Misra \& Sutaria (1999)
showed that this model fits the narrow band ASCA data for this source.
They had also pointed out that the broad band simultaneous data
obtained by BeppoSAX may rule out or confirm the model. Fitting this model
to the BeppoSAX data we obtain a  reduced
$\chi^2 = 136$ for 121 degrees of freedom. Since this is significantly
worse than the standard disk model, the Comptonization model
can be formally rejected thereby confirming that the line broadening
is due to gravitational effects.

\section{Summary and Discussion}

In this paper, we show that a double zone model for the
broad Iron line in AGN fits the ASCA and BeppoSAX
data for MCG-6-30-15. In this model, the X-ray source
is located around $\approx 10 r_g$ of the accretion disk
where there is maximum gravitational energy dissipation.
The line emission arises from a innermost disk region 
($\approx 7 r_g$) and from a region outside the X-ray source
(i.e. with radii $\approx 15 r_g$). The ASCA data reveals
that for this model the inclination angle of the source
$i \approx 10^o$ which is compatible with that obtained
by modeling of optical line emission from the same source
(Sulentic et al. 1998).

For several AGN the centroid of the blue wing is $\approx 6.4$ keV
even though the width of the line varies among the sources
(Nandra et al. 1997; Sulentic, Marziani \& Calvani 1998). 
The double zone model may be able to explain these observations,
if for most AGN the outer zone has a larger inner radius (i.e.
$R_{i2}$ is generally larger than what is obtained for
MCG-6-30-15) and the outer region is not highly ionized i.e.
the rest frame energy of the line photon is 6.4 keV. Note
the in the two zone model the width of the combined line
depends on the flux from the inner region. Hence the
blue and the red parts of the profile may vary independently
of each other. However, this speculation can only be
confirmed after detailed spectral fits by the double zone
model for several AGN is undertaken.

In the framework of the double zone model, the variability 
of MCG-6-20-15 can be explained by variation in $R_{i2}$
,$R_{o1}$ and/or the ionization state of the cold disk. 
Variations of $R_{i2}$ and $R_{o1}$ reflect the changing size 
of the X-ray emitting region with intensity. However,
the data is not statistically good enough to give any concrete
trends. 

Although the spectral shapes for the Iron line from the
standard disk model and the double zone one are similar
it may be possible to differentiate them by high resolution
future spectroscopy by satellites like XMM. One may also be
able to rule out or confirm this model by obtaining more 
concrete and independent estimation of the inclination angle
of MCG-6-30-15.

\acknowledgements

The author would like to thank Max Calvani and Mateo Guainazzi
for useful discussions and for making available the BeppoSAX data. The author
would also like to thank 
F. Sutaria for help with the X-ray data analysis.
 This research has made use of data obtained from the High Energy Astrophysics Science
Archive Research Center (HEASARC), provided by NASA's Goddard Space Flight Center.

\clearpage

\figcaption{ The line profile for the ASCA medium intensity (MI) data
set. The dashed lines are the profiles from the inner and outer regions.
The parameters for the fit are from Table 1: column 5\label{fig1}}

\figcaption{The line profile for the ASCA low intensity (LI) data
set. The dashed lines are the profiles from the inner and outer regions.
The parameters for the fit are from Table 3: column 5 \label{fig2}}

\clearpage

\plotone{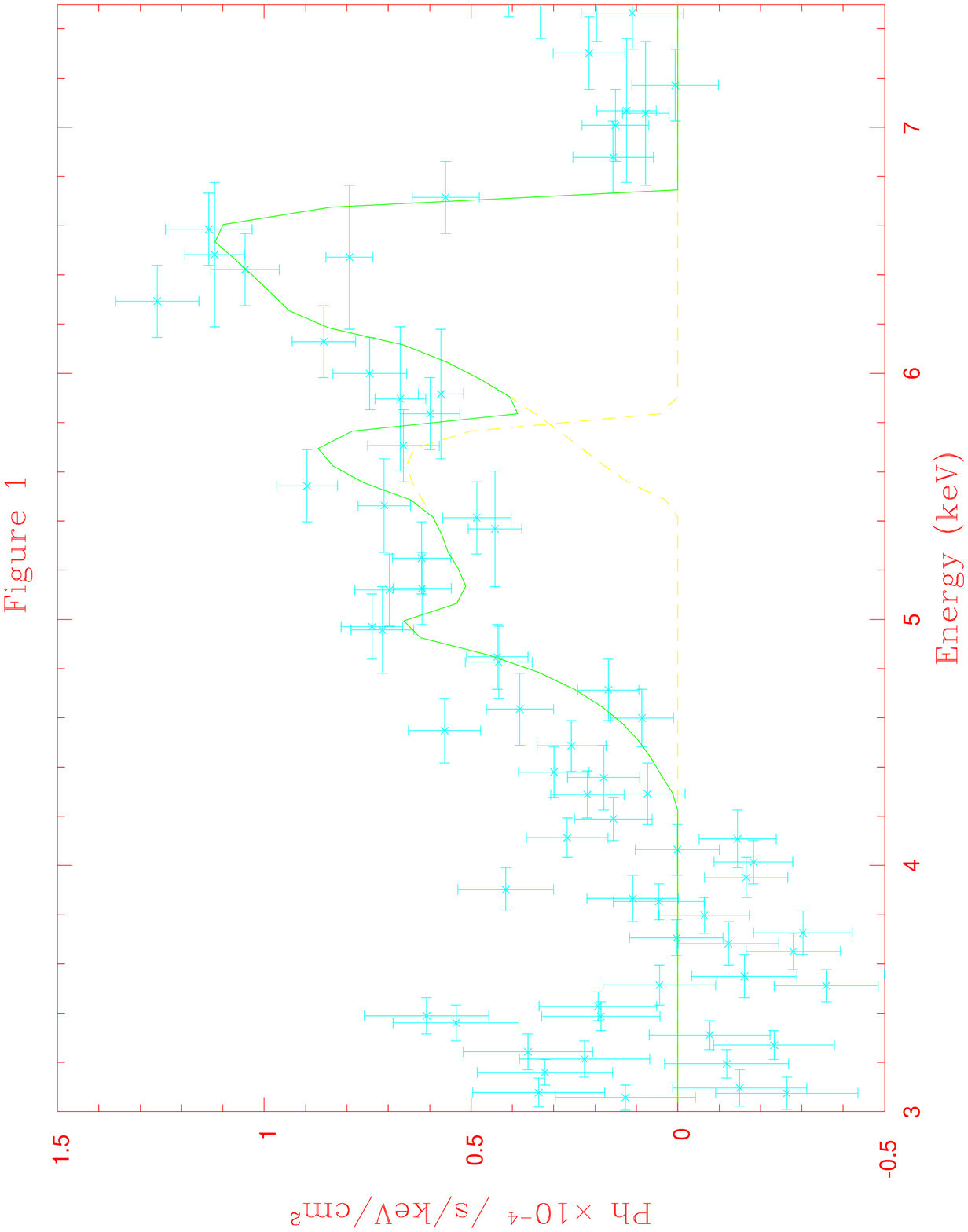}

\clearpage

\plotone{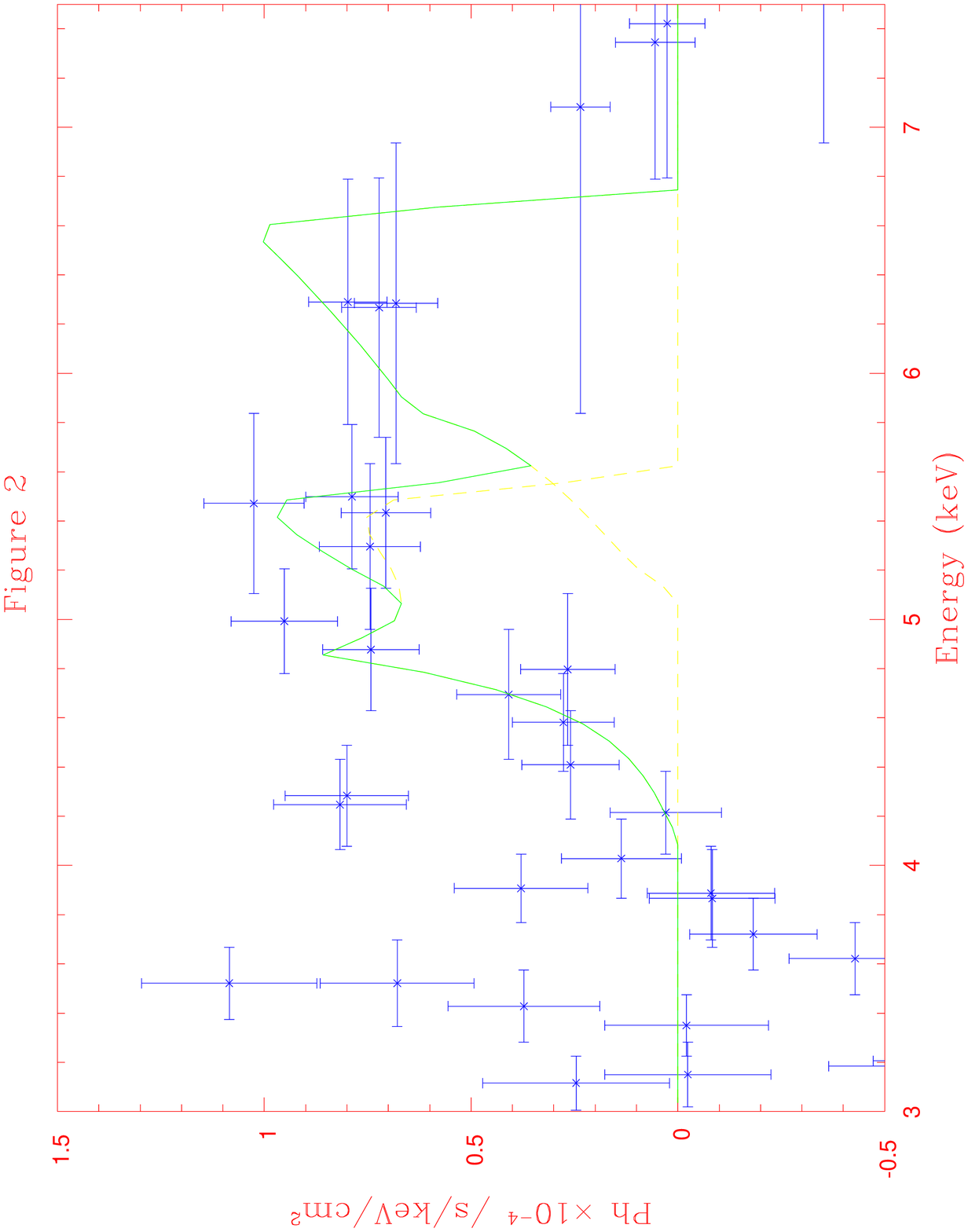}

\end{document}